\documentclass[preprint,review,12pt]{elsarticle}

\usepackage{epsfig}
\usepackage{multirow}
\usepackage{amssymb}

 \usepackage{lineno}

\journal{Journal of Nuclear Materials}

\begin{document}

\begin{frontmatter}



\title{First principle study of hydrogen behavior in hexagonal tungsten carbide}


\author{Xiang-Shan Kong$^{a}$, Yu-Wei You$^{a}$, C. S. Liu$^{a,\ast}\footnotetext{
*Author to whom correspondence should be
addressed. Email address: csliu@issp.ac.cn Tel: 0086-551-5591062}$,
Q. F. Fang$^{a}$, Jun-Ling Chen$^{b}$, G.-N. Luo$^{b}$,}


\address{$^{a}$Key Laboratory of Materials Physics, Institute of Solid
State Physics, Chinese Academy of Sciences, P. O. Box 1129, Hefei
230031, P. R. China

$^{b}$Institute of Plasma Physics, Chinese Academy of Sciences,
Hefei 230031, P. R. China}

\begin{abstract}
Understanding the behavior of hydrogen in hexagonal tungsten carbide
(WC) is of particular interest for fusion reactor design due to the
presence of WC in the divertor of fusion reactors. Therefore, we use
first-principles calculations to study the hydrogen behavior in WC.
The most stable interstitial site for the hydrogen atom is the
projection of the octahedral interstitial site on tungsten basal
plane, followed by the site near the projection of the octahedral
interstitial site on carbon basal plane. The binding energy between
two interstitial hydrogen atoms is negative, suggesting that
hydrogen itself is not capable of trapping other hydrogen atoms to
form a hydrogen molecule. The calculated results on the interaction
between hydrogen and vacancy indicate that the hydrogen atom is
energetically trapped by vacancy and the hydrogen molecule can not
be formed in mono-vacancy. In addition, the hydrogen atom bound to
carbon is only found in tungsten vacancy. We also study the
migrations of hydrogen in WC and find that the interstitial hydrogen
atom prefers to diffusion along the \textbf{\emph{c}} axis. Our
studies on the hydrogen behavior in WC provide some explanations for
the experimental results of the thermal desorption process of
energetic hydrogen ion implanted into WC.

\end{abstract}


\end{frontmatter}


\section{Introduction}
The present design for ITER foresees the use of three plasma facing
materials: Be for the first wall, W for the divertor and baffle
region and CFC for the divertor strike point
tiles.\cite{Kaufmann,Pamela,Doerner} The choice of the different
materials leads inexorably to the formation of mixed-materials
surface.\cite{Loarte} The mixed-materials can change not only the
thermo-mechanical properties, but also fuel retention properties of
the plasma facing wall, which influence the hydrogen recycling on
the plasma facing surface and the tritium inventory in the vacuum
vessel. In ITER, for safety reasons, periodic tritium removal will
be required before the in-vessel tritium inventory reaches its
administrative limit\cite{Federici}, meaning that tritium retention
rate strongly affects ITER operation program in the D-T phase. From
these considerations, hydrogen uptake and retention in mixed
material systems is an important issue for reliable extrapolation of
in-vessel tritium retention in ITER. Previous studies have indicated
that a WC layer is probably formed on tungsten surface succeeding
carbon impurity deposition.\cite{Eckstein,Kimura,Sugiyama,Ueda} Ueda
\emph{et al.}\cite{Ueda} found that all the tungsten atoms near the
top surface were combined with carbon atoms to form a tungsten
carbide layer, which prevented the implanted H from leaving the
tungsten and led to blister formation. Hence, as a fundamental
study, understanding the hydrogen behavior in WC is of particular
interest for fusion reactor design.

In the last decade, to understand the hydrogen behavior in WC, the
thermal desorption processes of energetic hydrogen or deuterium ion
implanted into WC have been studied by means of the thermal
desorption spectroscopy
(TDS).\cite{Horikawa1,Horikawa2,Kimura,Igarashi} Four desorption
peaks are observed, which are denoted by Peak 1, Peak 2, Peak 3, and
Peak 4, respectively, and described as follows: Peak 1 and Peak 2
observed at the lower temperature region are attributed to the
desorption of hydrogen retained in two different interstitial sites,
Peak 3 is due to the desorption of hydrogen trapped by carbon
vacancy, and Peak 4 is originated from the detrapping of hydrogen
bound to carbon. It should be pointed out that the hydrogen
concentration bound to carbon is very low and saturated at the
initial stage of hydrogen implantation and no hydrogen molecule
formation is observed in WC.\cite{Ogorodnikova} In addition,
Tr\"{a}skelin et al., \cite{Traskelin} have used the molecular
dynamics simulations to study the hydrogen bombardment of WC
surfaces. Even though certain details of the hydrogen trapping in WC
have been addressed by experimental and theoretical studies, several
questions on the trapping site of hydrogen and the interaction of
hydrogen with the carbon or tungsten vacancy in WC are still not
clearly understood. Besides, to our best knowledge, the physical
mechanism of hydrogen migration in WC has not been investigated
until now. In our preceding works \cite{Kong}, we have theoretically
investigated the properties of intrinsic defects in WC and revealed
the carbon vacancy to be the dominant defect in WC.

In this paper we employ first-principles calculations to study the
hydrogen behavior and the interaction of hydrogen with vacancy
defect in WC. The stable interstitial sites for hydrogen atom are
identified, and the interaction of double interstitial hydrogen
atoms in WC are investigated to understand whether hydrogen atom can
be self-trapping or forms a hydrogen molecule directly. The
processes of multiple hydrogen absorption into the vacancy are
studied to check the possibility for hydrogen occupancy of the
vacancy in WC. We also studied the migration of interstitial
hydrogen atom in WC and the detrapping of hydrogen from vacancy. Our
calculations are expected to reveal the physical mechanism of
hydrogen behavior in WC and provide some explanations for the
experimental results of the thermal desorption process of energetic
hydrogen ion implanted into WC.

\section{Computation method}

The present calculations have been performed within density
functional theory as implemented in the Vienna \emph{ab} initio
simulation package (VASP) \cite{Kresse1,Kresse2}. The interaction
between ions and electrons is described by the projector augmented
wave potential (PAW) method \cite{Kresse3,Blohl}. Exchange and
correlation functions are taken in a form proposed by Perdew and
Wang (PW91) \cite{Perdew} within the generalized gradient
approximation (GGA). The supercell approach with periodic boundary
conditions is used. Two types of hexagonal simulation supercells are
used: the smaller one contains 54 atoms (27 carbon atoms and 27
tungsten atoms) with $3\times3\times3$ unit cells; the larger one
contains 128 atoms (64 carbon atoms and 64 tungsten atoms) with
$4\times4\times4$ unit cells. The grids of \emph{k}-points centered
at the Gamma point are sampled by $5\times5\times5$ for 54-atom
supercell and $3\times3\times3$ for the 128-atom supercell. The
relaxations of atomic position and optimizations of the shape and
size of the supecell are performed with the plane-wave basis sets
with the energy cutoff of 500 eV throughout this work, which was
checked for convergence to be within 0.001 eV per atom in the
perfect supercell. The structural optimization is truncated when the
forces converge to less than 0.1 eV/nm. In addition, the diffusion
properties of hydrogen in WC are calculated by the nudged elastic
band method with the 54-atom supercell.

The ground-state properties of WC, including equilibrium lattice
parameters and bulk cohesive energy, have been calculated in order
to compare with experimental and former theoretical data. Results
are presented in Table 1. In addition, we obtained the bulk modulus
by fitting the volume and calculated cohesive energy to Murnaghan's
equation of the state. It can be seen from Table 1 that our results
are in agreement with the available experimental and theoretical
values.

Firstly, to identify the energy favorable trapping site, the
formation energies of single hydrogen atom occupying different
interstitial sites in WC are calculated by
\begin{equation}
E_{f}=E_{tot}^{H}-E_{tot}-\mu_{H},
\end{equation}
where $E_{tot}$ is the total energy of WC, $E_{tot}^{H}$ is the
total energy of the system with just a hydrogen atom and $\mu_{H}$
is the hydrogen chemical potential. And then, we use the binding
energy to investigate the interaction of double interstitial
hydrogen atoms in WC, which is defined as
\begin{equation}
E_{b}=2E_{tot}^{H}-E_{tot}^{2H}-E_{tot},
\end{equation}
where $E_{tot}^{2H}$ is the total energy of the WC with two
interstitial hydrogen atoms. Here, negative bonding energy indicates
repulsion between hydrogen atoms, while positive binding energy
means attraction. Lastly, the interaction between hydrogen and
carbon or tungsten vacancy is studied by estimation of the trapping
energy. For clarity, below we illustrate the calculated formula only
in the case of carbon vacancy, but the calculation formula can be
simply applied for tungsten vacancy by replacing carbon by tungsten.
Cumulative hydrogen trapping energy of \emph{m} hydrogen atoms in a
carbon vacancy is calculated by using the following expression
\cite{Yu}:
\begin{equation}
E_{tr}^{mH-V_{C}}=E_{tot}^{mH-V_{C}}-E_{tot}^{V_{W}}-m(E_{tot}^{H}-E_{tot}),
\end{equation}
where \emph{m} is the number of hydrogen atoms and
$E_{tot}^{mH-V_{C}}$ is the total energy of WC with $m$ hydrogen
atoms and a carbon vacancy. Accordingly, the trapping energy $\Delta
E_{tr}^{mH-V_{C}}$ when the number of hydrogen atoms is increased
from $m-1$ to $m$ is defined as:
\begin{equation}
\Delta
E_{tr}^{mH-V_{C}}=E_{tot}^{mH-V_{C}}-E_{tot}^{(m-1)H-V_{C}}-(E_{tot}^{H}-E_{tot}).
\end{equation}

\section{Results and discussion}

\subsection{Single hydrogen atom in WC}

There are ten possible interstitial sites for hydrogen in WC, as
shown in Fig. 1. which are denoted by the symbols O, BOC, BOW, TC,
TW, BTC, BTW, C, BCC, and BCW, and described as follows. O is the
octahedral interstitial sites formed by equivalent three tungsten
atoms and three carbon atoms. BOC and BOW are projections of O site
on carbon and tungsten basal plane, respectively. TC and TW are two
different tetrahedral interstitial sites formed by two different
atomic clusters: the former is composed of three carbon atoms and
one tungsten atom and the later consists of three tungsten atoms and
one carbon atom. Similarly, BTC and BTW are two different hexahedral
sites formed by two different atomic clusters: the former is formed
by three carbon atoms and two tungsten atoms, while the later is
made up of three tungsten atoms and two carbon atoms. C is midway
between the nearest-neighbor tungsten atom and carbon atom. BCC and
BCW are the midway between two nearest carbon and tungsten atoms in
the dense \textbf{\emph{a}} direction, respectively. It should be
noted that BTC and BTW are also the projection of TC and TW on the
carbon and tungsten basal plane, respectively. All configurations
with a hydrogen atom in interstitial site are fully relaxed, and two
stable hydrogen interstitial configurations are found: one is that
the hydrogen atom occupies the BOW site with a formation energy of
-0.54 eV, the other is that the hydrogen atom is located in the site
near BOC (namely NBOC, Fig. 1) with a formation energy of 0.07 eV.
The hydrogen detrapping from BTW and NBOC sites may be responsible
for the Peak 2 and Peak 1, respectively. The present results about
interstitial site of hydrogen are disagreement with the conclusion
of Horikawa and Igarashi \cite{Horikawa2,Igarashi} that the stable
interstitial sites for hydrogen in WC is BTW and BTC. Both of them
did not consider other interstitial sites in WC except BTW and BTC.
In order to study the effect of supercell size on the stability of
hydrogen interstitial configurations, similar calculations have been
performed in 54-atom and 128-atom supercell. It is interest to find
that changing the supercell size from 128 to 54 slightly affects the
formation energies of the hydrogen interstitial, but it does not
change final configurations and their relative stability. The
relative energy differences between the defect configurations change
by at most 0.06 eV. In the following we report only the results
obtained from 128-atom supercell calculations, unless clearly stated
otherwise.

\subsection{Double interstitial hydrogen atoms in WC}

We now investigate the interaction of double interstitial hydrogen
atoms in WC, in order to understand whether hydrogen atom can be
self-trapping or forms a hydrogen molecule directly. In the WC,
there are distinct sites aligned along the \emph{\textbf{c}}
directions and in the carbon and tungsten basal plane. Hence,
according to the above calculated results, we investigate the
interaction of interstitial hydrogen atoms in the following cases:
(i) the first hydrogen atoms is placed in the BOW site, and the
second atom is inserted at other interstitial site in the same
tungsten basal plane; (ii) the first hydrogen atom sits in  the NBOC
site, and the second atom occupies other interstitial site in the
same carbon basal plane; (iii) two interstitial hydrogen atoms along
the \emph{\textbf{c}} axis. All relative initial configurations with
double interstitial hydrogen atoms are fully relaxed, and eleven
final stable configurations are obtained as shown in Fig. 3. The
binding energies of these eleven final configurations are calculated
using the Eq. 2. The initial and finial distance between two
hydrogen atoms and the corresponding binding energies are summarized
in Table 2. As shown in Table 2, the distance between two hydrogen
atoms increases, and the binding energy are negative in all
configurations. They indicate the present of a repulsive interaction
when interstitial hydrogen atom is close to each other. It should be
noted that the smallest distance between two hydrogen atoms is 0.144
nm, much longer than the bond length of 0.075 nm in hydrogen
molecular. In a word, the present results suggest that hydrogen
atoms cannot bind together to form a hydrogen molecule in the
perfect bulk WC, and thus no hydrogen bubbles.

\subsection{Hydrogen-vacancy interaction in WC}

In order to check the possibility for hydrogen occupancy of the
vacancy in WC, which has been predicted experimently and in
molecular dynamic simulation
\cite{Horikawa1,Horikawa2,Kimura,Igarashi,Traskelin}, we study in
this section multiple hydrogen absorption process into the vacancy.
In WC, there are two types of vacancy, namely carbon and tungsten
vacancy. For the carbon vacancy, there are six equivalent tungsten
atoms around the vacancy site as its first-neighbors with a distance
of 0.221 nm and six carbon atoms in the carbon basal plane as its
second-neighbors with a distance of 0.293 nm. Similar neighbors
exist around the tungsten vacancy site: six carbon atom as the
first-neighbors and six tungsten atom as the second-neighbors.
Therefore, we investigation hydrogen occupancy in carbon and
tungsten vacancy, respectively. For clarity, below we illustrate the
simulation method only in the case of carbon vacancy, and the
similar processes are used in the case of tungsten vacancy. We first
examine several possible occupied sites of hydrogen in carbon
vacancy and find the most stable site according to the calculated
total energy. And then, we bring the hydrogen atom one by one to the
vacancy and minimize the energy to find the most stable
configurations of $mH-V_{C}$ complex. In each step, we investigate
up to twenty possible configurations based on the most stable
configurations of $(m-1)H-V_{C}$ complex.

The stable configurations for $mH-V_{C}$ complexes are represented
in Fig. 3, and the corresponding trapping energy for the $m$-th
hydrogen atom by the $(m-1)H-V_{C}$ cluster ($\Delta
E_{tr}^{mH-V_{C}}$) as well as the cumulative trapping energy for
trapping $m$ hydrogen atoms in a carbon vacancy
($E_{tot}^{mH-V_{C}}$), calculated by the Eqs. (3) and (4), are
listed in Table 3. As shown in Fig. 4 (a) and (e), the most stable
site for single hydrogen atom is to be at an off-vacancy-center
position (0.047 nm from the vacancy center). In the $mH-V_{C} (m>1)$
complexes, the hydrogen atom prefers to occupy the projection sites
of the BCW on the carbon basal plane and forms four W-H bonds with
the nearest tungsten atoms with bond length of about 0.21 nm. It
should be noted that the hydrogen behavior in carbon vacancy in WC
are similar with that in tungsten vacancy in tungsten \cite{Lu}.
This may be one reason that the desorption temperature of Peak 3 are
nearly consistent with the temperature of the desorption peak
attributed to the desorption of hydrogen trapped by tungsten
vacancies in tungsten \cite{Kimura}. As shown in Table 3, the
trapping energies of $mH-V_{C}$ complexes are negative, and the
absolute values of the cumulative trapping energy increases with the
number of hydrogen atoms until $m = 3$. This indicates that the
addition of hydrogen atoms is energetically favorable. For the
$4H-V_{C}$ complex, the trapping energy is positive, and the
cumulative energy become less negative than $3H-V_{C}$. Thus, we
make the conclusion that it is energetically favorable for a
mono-carbon-vacancy to trap as many as 3 hydrogen atoms. It should
be pointed that the lowest distance between two hydrogen atoms in
stable $mH-V_{C}$ complexes is 0.166 nm, much longer than the
 bond length of hydrogen molecule, implying that hydrogen atoms cannot bind
together to form a hydrogen molecule in mono-carbon-vacancy.

Figure 4 represents the most stable configurations of $mH-V_{W}$
complexes obtained following the aforementioned method, and Table 4
list the corresponding trapping energy and cumulative trapping
energy. In tungsten vacancy, a single hydrogen atom prefers to
occupy the interstitial C site to form a C-H bond with the nearest
carbon atom with the bond length of 0.116 nm. In $mH-V_{W} (m>1)$
complexes, the hydrogen atoms sit in the C and BOW around the vacant
site, and detailed structural properties are summarized in Table 4,
such as the occupancy sites of $m$ hydrogen atoms, the number and
length of the C-H bonds, and the smallest distance between two
hydrogen atoms in $mH-V_{W}$ cluster. As shown in Table 4, the
trapping energy varies slowly at first but then very rapidly becomes
less negative with the increasing the number of hydrogen atoms up to
9, beyond which the trapping energy becomes positive. In addition,
the cumulative trapping energy increases in its absolute value with
the number of hydrogen atoms in $mH_{W}$ cluster until $m=9$. Thus,
we suggest that the maximum number of hydrogen atoms are trapped by
mono-tungsten-vacancy is 9. Note that, in all stable $mH-V_{W}$
clusters, the H-H distance are much larger than the bond length of
$H_{2}$, indicating that the hydrogen molecule can not be formed in
mono-tungsten-vacancy. In addition, there are as many as 6 C-H bonds
in a tungsten vacancy. It should be pointed out that the hydrogen
atoms bound to carbon are only found in tungsten vacancy whose
concentration in WC is very small. This is why the concentration of
hydrogen bound to carbon in WC is very low \cite{Kimura}.

\subsection{Migration of interstitial hydrogen atom in WC}

Concerning the kinetics of hydrogen in WC, we first examine the case
of interstitial hydrogen migration, which is relevant to the initial
stage after hydrogen implantation or low temperature re-emission of
hydrogen atoms trapped in WC. As shown in Fig. 5, we investigate
four routes named by Path A, Path B, Path C and Path D,
respectively. Path A ($4\rightarrow5$) is the migrate of hydrogen
atom between two nearest equivalent BOW site with the energy barrier
of 1.79 eV; Path B ($1\rightarrow2$) and Path C ($2\rightarrow3$)
are the diffusions of hydrogen atom in carbon basal plane: the
former is the diffusion of hydrogen atom between two
nearest-neighbor NBOC sites around the same BOC site with a barrier
energy of 0.23 eV, while the later is that around different BOC site
with a large energy barrier of 0.84 eV. Path D ($1\rightarrow4$)
propagates parallel to the \emph{\textbf{c}} direction, and the
energy barrier of hydrogen jumping from the NBOC to the nearest BOW
is 0.05 eV, while the reverse diffusion energy barrier is 0.5 eV.
Therefore, the interstitial hydrogen atom prefers to jump form NBOC
to BOW along the \emph{\textbf{c}} axis in WC.

We now turn to the kinetic process for hydrogen detrapping from
carbon vacancy in WC. Similarly, we investigate two diffusion routes
for hydrogen detrapping from carbon vacancy shown in Fig. 6: one is
that hydrogen atom diffusion in the carbon plane ($1\rightarrow2$,
Path E); the other is parallel to the \emph{\textbf{c}} axis
($1\rightarrow3\rightarrow4$, Path F). For path E, the energy
barrier for hydrogen atoms jumping away from the carbon vacancy is
2.1 eV, which are greatly lager than the corresponding reverse
energy barrier (0.73 eV). Path F involves two steps for hydrogen
detrapping from carbon vacancy. The first step is that hydrogen atom
jump form site 1 to 3 with a diffusion barrier of 0.64 eV, and the
second step is hydrogen diffuse from  site 3 to 4 with a energy
barrier of 1.06 eV. Not that, in both Path E and Path F, the
hydrogen detrapping energy barrier are much larger than hydrogen
tapping energy barrier. This means that the hydrogen atom is
energetically trapped by the carbon vacancy. It should be pointed
out that the energy barrier of interstitial hydrogen diffusion near
the carbon vacancy becomes smaller than that in the area without
carbon vacancy.

As similar as the case of hydrogen detrapping from carbon vacancy,
we study two diffusion routes for hydrogen detrapping from tungsten
vacancy shown in Fig. 7: one is that hydrogen diffusion in the
tungsten basal plane ($1\rightarrow2\rightarrow3$, Path G); the
other propagates along the \emph{\textbf{c}} axis
($1\rightarrow2\rightarrow4$, Path J). Both of them involve the step
that hydrogen atom jumps from C site to the BOW in tungsten vacancy
with the energy barrier of 0.58 eV. For Path G, the energy barrier
of hydrogen atoms jumping away from the tungsten vacancy is 3.88 eV,
greatly larger than that along the \emph{\textbf{c}} axis. It should
be noted that the energy barrier of hydrogen atom jumping into the
tungsten vacancy is relatively low. Together with the low
concentration of hydrogen bound to carbon vacancy, may provide an
explanation for that the concentration of hydrogen bound to carbon
is saturated at the initial stage of hydrogen implantation
\cite{Kimura}.

\section{conclusion}

We use first-principles calculations to study the hydrogen behavior
in WC. Two stable interstitial sites for hydrogen atoms, namely BOW
and NBOC. BOW is more stable for interstitial hydrogen atoms than
NBOC. The binding energy between two interstitial hydrogen atoms is
negative, suggesting that hydrogen itself is not capable of trapping
other hydrogen atoms to form a hydrogen molecule. The calculated
results on the interaction between hydrogen and vacancy indicate
that the maximum number of hydrogen atoms trapped by
mono-carbon-vacancy and mono-tungsten-vacancy are 3 and 9,
respectively, and the hydrogen molecule can not be formed in
mono-carbon-vacancy and mono-tungsten-vacancy. It should be noted
that the hydrogen atoms bound to carbon are only found in tungsten
vacancy whose concentration in WC is very small. This is why the
concentration of hydrogen bound to carbon in WC is very low. We also
investigate the migration of hydrogen in WC with and without vacancy
defect. The calculated results suggest that the interstitial
hydrogen atom prefers to diffusion along the \textbf{\emph{c}} axis,
and the hydrogen atom are easily diffusion into and trapped by the
vacancy. It should be pointed out the energy barrier of the hydrogen
atom jumping into the tungsten vacancy is very low. This provides an
explanation for that the concentration of hydrogen bound to carbon
is saturated at the initial stage of hydrogen implantation in WC.

\section*{Acknowledgement}

This work was supported by the National Magnetic Confinement Fusion
Program (Grant No.: 2009GB106005) and the Innovation Program of
Chinese Academy of Sciences (Grant No.: KJCX2-YW-N35), and by the
Center for Computation Science, Hefei Institutes of Physical
Sciences.




\bibliographystyle{elsarticle-num}
\bibliography{<your-bib-database>}






\newpage
\begin{center}
Table 1 The bulk properties and cohesive energy ($E_{c}$) of WC, MD
and DFT mean the results obtained by molecular dynamic simulation
and density functional theory, respectively.

\mathstrut

\begin{tabular}{ccccc}
\hline \hline & $a_{0}($nm$)$ & $c_{0}/a_{0}$ & $B_{0}($GPa$)$ &
$E_{c}($eV$)$
\\ \hline
\multicolumn{1}{l}{Experiment} & 0.2907$^{a}$ & 0.976$^{a}$ & 329$^{b}$ & -16.68$^{c}$ \\
\multicolumn{1}{l}{MD$^{d}$} & 0.2917 & 0.964 & 443 & -16.68 \\
\multicolumn{1}{l}{DFT$^{d}$} & 0.2979 & 0.975 & 368 & -15.01 \\
\multicolumn{1}{l}{Present work} & 0.2932 & 0.973 & 356 & -16.42\\
\hline\hline\\
$^{a}$ Reference \cite{Pierson}\\
$^{b}$ Reference \cite{Brown}\\
$^{c}$ Reference \cite{CRC}\\
$^{d}$ Reference \cite{Juslin}\\
\end{tabular}

\newpage

Table 2 The initial and final distance (before and after the
relaxation, nm) between two hydrogen atoms and corresponding binding
energy (eV) in three cases.

\mathstrut

\begin{tabular}{cccccc}
\hline\hline case & Site of second atom & Initial distance & Final
distance & Bonding energy \\ \hline
(i) & 1 & 0.1681 & 0.1743 & -1.92 \\
& 2 & 0.2906 & 0.2971 & -0.18 \\
& 3 & 0.5033 & 0.5093 & -0.06 \\
& 4 & 0.5812 & 0.5875 & -0.06 \\
(ii)& 1 & 0.1163 & 0.1444 & -0.59 \\
& 2 & 0.2932 & 0.2983 & -0.11 \\
& 3 & 0.3467 & 0.3503 & -0.09 \\
& 4 & 0.3849 & 0.3856 & -0.07 \\
& 5 & 0.5865 & 0.5866 & -0.05 \\
(iii)& 1 & 0.2838 & 0.2920 & -0.17 \\
& 2 & 0.5674 & 0.5721 & -0.02\\ \hline\hline
\end{tabular}

\newpage
Table 3 The trapping energy for the $m$-th hydrogen atom by the
$(m-1)H-V_{c}$ cluster ($\Delta E_{tr}^{mH-V_{C}}$, eV) and the
cumulative trapping energy for trapping $m$ hydrogen atoms in a
carbon vacancy ($E_{tot}^{mH-V_{C}}$, eV).

\mathstrut

\begin{tabular}{ccc}
\hline\hline system & $\Delta
E_{tr}^{mH-V_{C}}$ & $E_{tot}^{mH-V_{C}}$\\
\hline
$1H-V_{C}$ & -0.25 & -0.25 \\
$2H-V_{C}$ & -0.83 & -1.08 \\
$3H-V_{C}$ & -0.27 & -1.35 \\
$4H-V_{C}$ & 0.20 & -1.15 \\
\hline\hline
\end{tabular}

\newpage
Table 4 The trapping energy for the $m$-th hydrogen atom by the
$(m-1)H-V_{W}$ cluster ($\Delta E_{tr}^{mH-V_{W}}$, eV) and the
cumulative trapping energy for trapping $m$ hydrogen atoms in a
tungsten vacancy ($E_{tot}^{mH-V_{W}}$, eV). $N_{C-H}$ and $D_{C-H}$
(nm) are the number and length of C-H bond, respectively. $D_{H-H}$
(nm) is the smallest distance between two hydrogen atoms.

\mathstrut

\begin{tabular}{ccccccc}
\hline\hline  system & Occupied site & $\Delta
E_{tr}^{mH-V_{W}}$ & $E_{tot}^{mH-V_{W}}$ & $N_{C-H}$ & $D_{C-H}$ & $D_{H-H}$\\
\hline
$1H-V_{W}$ & 1 & -0.95 & -0.95 & 1 & 0.116 &  \\
$2H-V_{W}$ & 1,2 & -0.89 & -1.85 &2 & 0.115 & 0.224  \\
$3H-V_{W}$ & 1$\sim$3 & -0.93 & -2.78 & 2 & 0.116 & 0.178 \\
$4H-V_{W}$ & 1$\sim$4 & -0.92 & -3.70 & 2 & 0.116 & 0.174 \\
$5H-V_{W}$ & 1$\sim$5 & -0.95 & -4.66 & 2 & 0.116 & 0.175 \\
$6H-V_{W}$ & 1$\sim$6 & -0.70 & -5.36 & 3 & 0.117 & 0.153\\
$7H-V_{W}$ & 1$\sim$7 & -0.62 & -5.97 & 4 & 0.114 & 0.155\\
$8H-V_{W}$ & 1$\sim$8 & -0.07 & -6.04 & 5 & 0.114 & 0.149\\
$9H-V_{W}$ & 1$\sim$9 & -0.01 & -6.05 & 6 & 0.111 & 0.149\\
$10H-V_{W}$ & 1$\sim$9,sub & 1.48 & -4.57 & 6 & 0.113 & 0.129  \\
\hline\hline
\end{tabular}

\end{center}


\newpage
\begin{figure}[h]
\begin{center}
\mbox{\epsfig{file=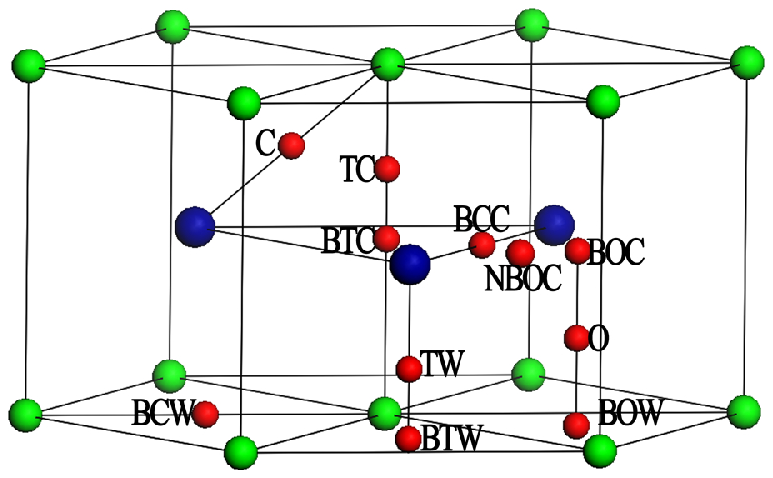}}
\end{center}
\caption{(color online)Schematic diagram of various isolated
interstitial sites in WC. The tungsten and carbon atoms are marked
in green (white) and blue (black) balls, respectively. The small red
(dark gray) balls represent the interstitial sites. }
\end{figure}
\clearpage

\newpage
\begin{figure}[h]
\begin{center}
\mbox{\epsfig{file=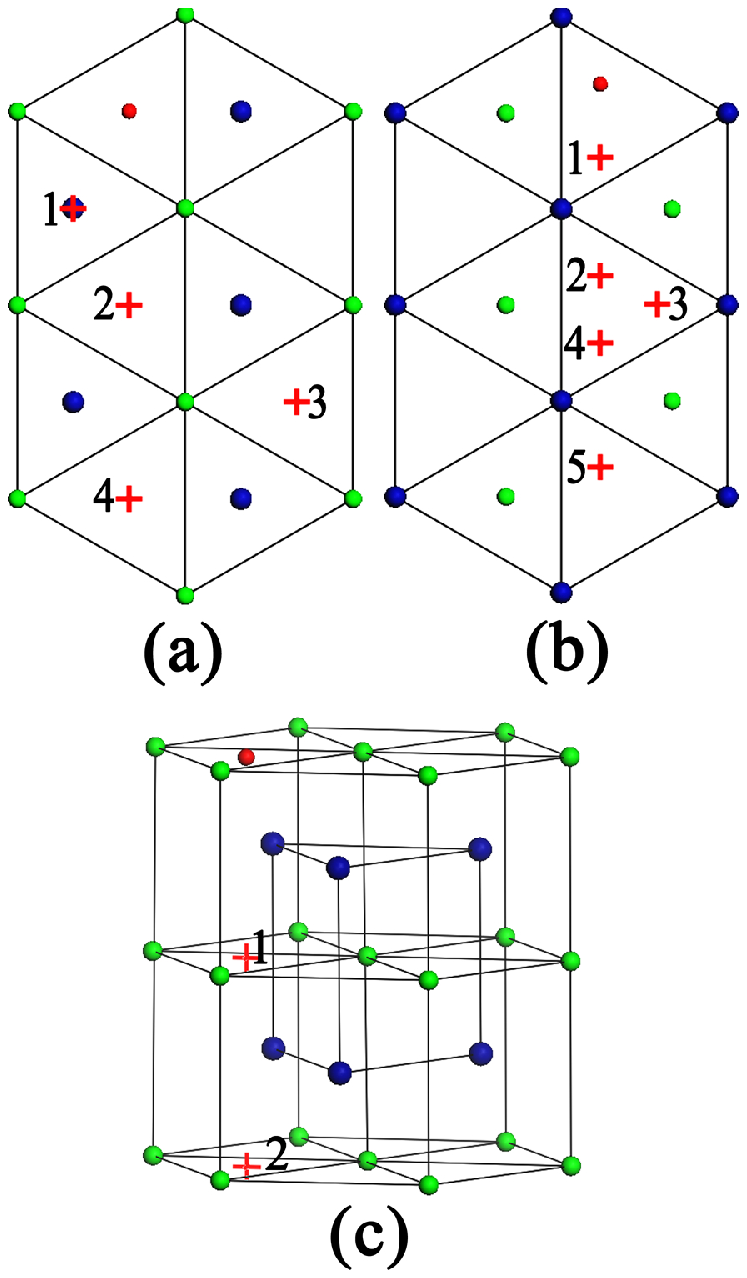}}
\end{center}
\caption{(color online) The final stable configurations of double
interstitial hydrogen atoms in WC. The (a), (b) and (c) show the
final configurations of case (i), (ii) and (iii), respectively. The
tungsten and carbon atoms are marked in green (white) and blue
(black) balls, respectively. The small red (dark gray) balls
represent the first interstitial hydrogen atom, and the red `+',
named by 1 $\sim$ 5, denotes the sites occupied by the second
hydrogen atom.}
\end{figure}
\clearpage

\newpage
\begin{figure}[h]
\begin{center}
\mbox{\epsfig{file=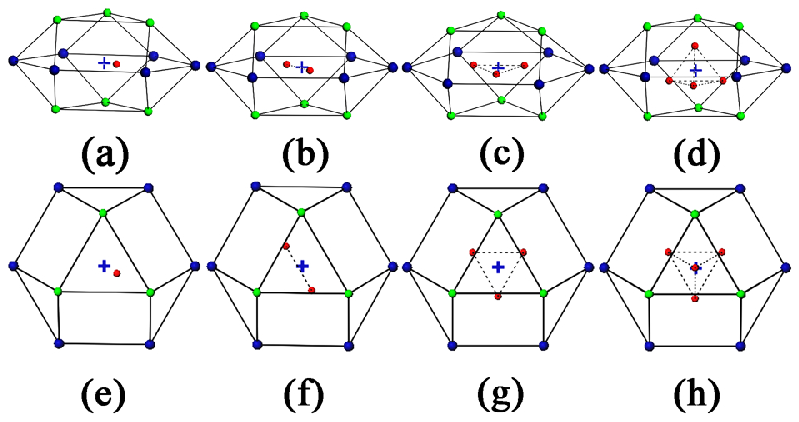}}
\end{center}
\caption{(color online) Schematic representation of the
lowest-energy configurations $mH-V_{C}$: (a) $1H-V_{C}$, (b)
$2H-V_{C}$, (c) $3H-V_{C}$, (d) $4H-V_{C}$. The (e)$\sim$(f) are
top-view images of (a)$\sim$(d), respectively. The tungsten, carbon
and hydrogen atoms are marked in green (white), blue (black), and
red (gray) balls, respectively. The green `+' represents the carbon
vacancy site.}
\end{figure}
\clearpage

\newpage
\begin{figure}[h]
\begin{center}
\mbox{\epsfig{file=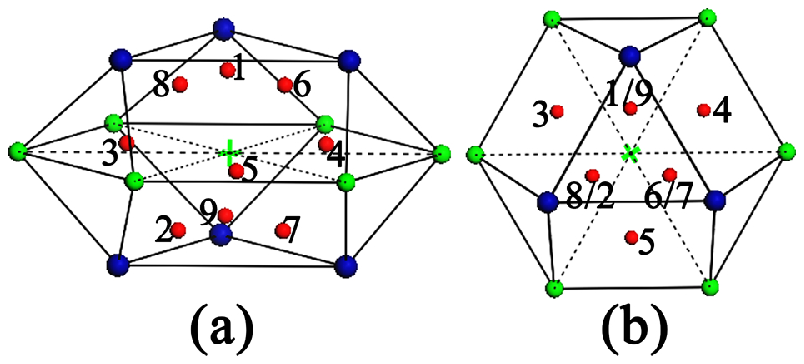}}
\end{center}
\caption{(color online) Schematic representation of the
lowest-energy configurations $mH-V_{W}$. The (b) is top-view images
of (a). The tungsten and carbon atoms are marked in green (white)
and blue (black), respectively. The blue `+' represents the tungsten
vacancy site. The sites occupied by the hydrogen atoms are marked by
the red balls, named by 1$\sim$9, respectively.}
\end{figure}
\clearpage

\newpage
\begin{figure}[h]
\begin{center}
\mbox{\epsfig{file=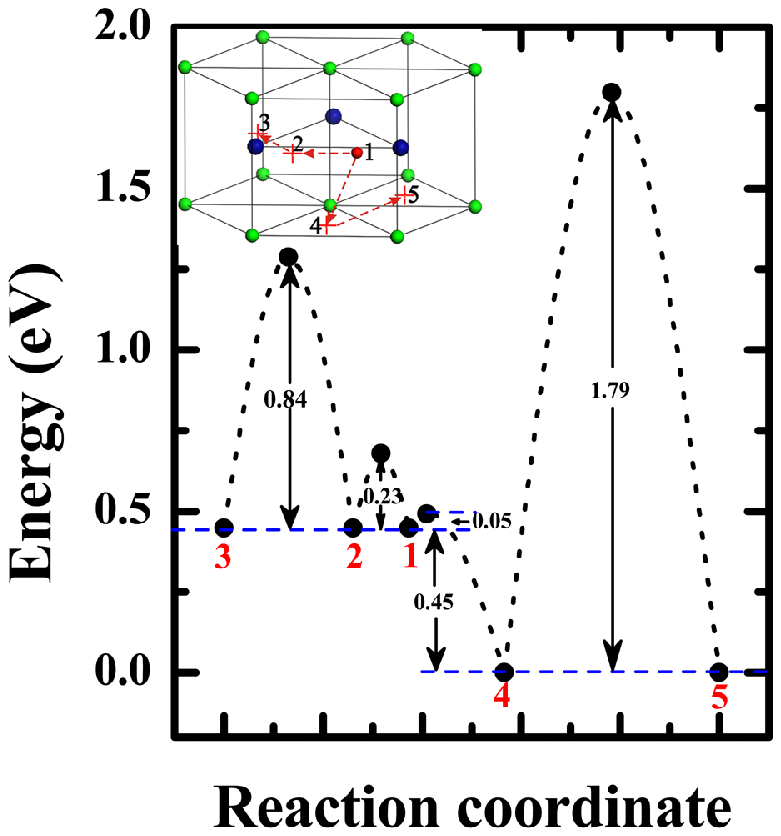}}
\end{center}
\caption{(color online) Diffusion energy profile and the
corresponding diffusion paths for interstitial hydrogen atom in WC.
The tungsten and carbon atoms are marked in green (white) and blue
(black), respectively. Site 1, 2 and 3 are the different NBOC sites
in a carbon basal plane and Sites 4 and 5 are two nearest BOW sites
in a tungsten basal plane.}
\end{figure}
\clearpage

\newpage
\begin{figure}[h]
\begin{center}
\mbox{\epsfig{file=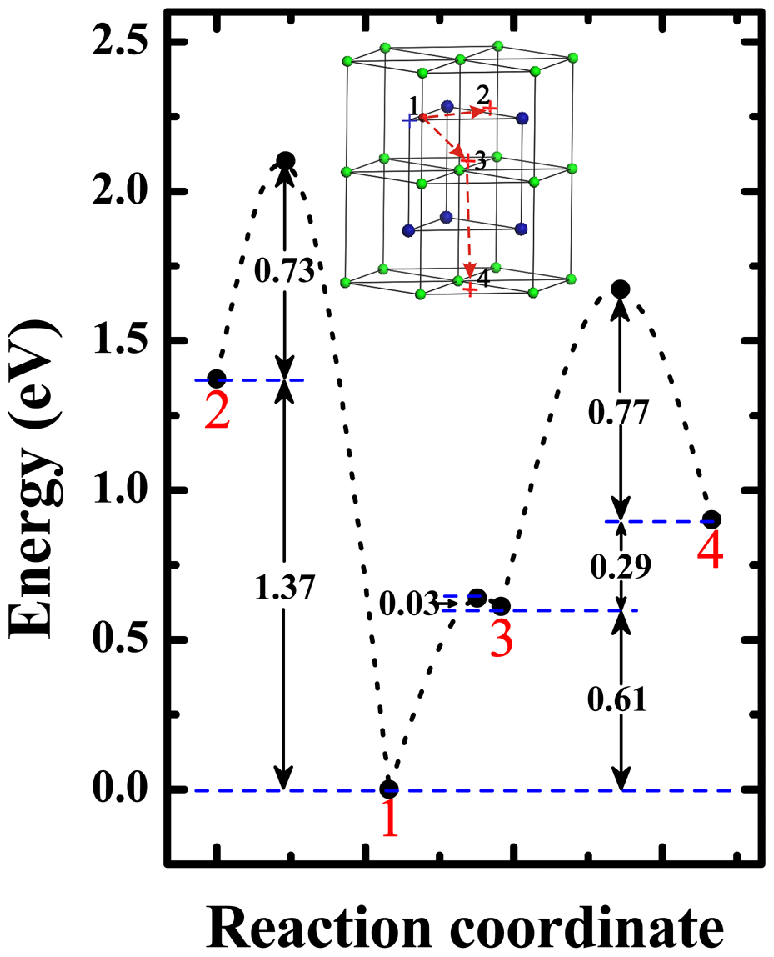}}
\end{center}
\caption{(color online) Hydrogen diffusion energy profile and the
corresponding diffusion paths around the carbon vacancy in WC. The
tungsten, carbon and hydrogen atoms are marked in green (white),
blue (black), and red (gray) ball, respectively. The green `+'
represents the carbon vacancy site. The red `+' are the relative
stable sites of hydrogen. Site 1 is the most stable site of hydrogen
in carbon vacancy. Sites 2 and 3 are the first nearest neighbor NBOC
and O site of the carbon vacancy, respectively. Site 4 is the second
nearest neighbor BOW site of the carbon vacancy.}
\end{figure}
\clearpage

\newpage
\begin{figure}[h]
\begin{center}
\mbox{\epsfig{file=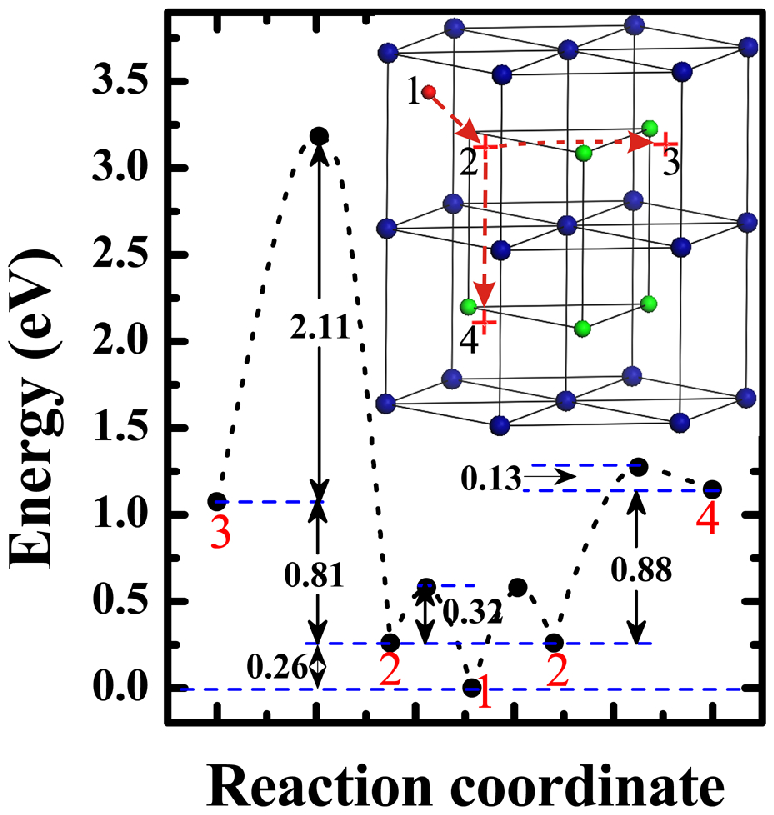}}
\end{center}
\caption{(color online) Hydrogen diffusion energy profile and the
corresponding diffusion paths around the tungsten vacancy in WC. The
tungsten, carbon and hydrogen atoms are marked in green (white),
blue (black), and red (gray) ball, respectively. The blue `+'
represents the tungsten vacancy site. The red `+' are the relative
stable sites of hydrogen. Site 1 is the most stable site of hydrogen
in tungsten vacancy. Site 2 is the BOW sites in tungsten vacancy.
Sites 3 and 4 are the nearest BOW site of site 2 along the c
direction and in same tungsten basal plane, respectively.}
\end{figure}
\clearpage

\end{document}